\documentclass{sanket}
\usepackage[colorlinks=true,citecolor=blue,linkcolor=blue]{hyperref}

\usepackage{amssymb}
\usepackage{graphicx}
\usepackage{subfigure}

\usepackage{amsmath}
\usepackage{xcolor}
\usepackage{mathrsfs}
\usepackage{indentfirst}
\usepackage{textcomp}
\usepackage{comment}
\usepackage{mathtools}

\newcommand{\ignore}[1]{}

\usepackage{natbib}

\allowdisplaybreaks

\newcommand{\tJmodel}{$t$-$t'$-$t''$-$J$~}

\begin{document}
\title{Data-efficient surrogate modeling of spectral functions using Gaussian processes: An application to the \texorpdfstring{$t$-$t'$-$t''$-$J$}{t-t'-t''-J} model}
\author[1]{Sanket Jantre}
\author[1]{Nathan M. Urban}
\author[2]{Weiguo Yin}
\author[2]{Niraj Aryal\thanks{corresponding author}}
\affil[1]{Applied Mathematics Department, Computing \& Data Sciences Directorate, Brookhaven National Laboratory}
\affil[2]{Condensed Matter Physics \& Materials Science Division, Brookhaven National Laboratory}
\date{}

\maketitle

\begin{abstract}
Spectral functions encode key many-body information but are costly to compute with high fidelity. Machine-learning surrogates have emerged as a powerful alternative, yet many approaches require large training datasets. We develop a data-efficient surrogate for spectral functions using the \tJmodel model, which describes the motion of a hole in a quantum antiferromagnet. Using $\sim$ 10$^5$ self-consistent Born approximation-based spectra from Lee, Carbone and Yin (Phys. Rev. B 107, 205132 (2023)), we train a deep-kernel Gaussian process surrogate model with sparse variational inference (DKL-SVGP) using only 10\% of the available training spectra. We benchmark against feed-forward neural networks (FFNN) trained on the same reduced subset and on the full dataset. The proposed DKL-SVGP model consistently outperforms the reduced-data FFNN and, despite using only 10\% of the training spectra, achieves spectrum-wise errors within the same order-of-magnitude as the full-data FFNN baseline. Worst-tail diagnostics show improved fidelity on difficult spectra, while peak-level analysis indicates that DKL-SVGP recovers dominant peak heights with comparable accuracy and improves peak-location agreement under a matched-peak evaluation that mitigates rare peak-swapping cases. Overall, these results highlight GP-based surrogates as a competitive and data-efficient approach for spectral-function prediction in scarce-data regimes. The DKL-SVGP model is implemented in the open source software Python and is available at: \url{https://github.com/jsanket12/Spectral_fn/}.
\end{abstract}

\maketitle

\section{Introduction}
Spectral functions sit at the heart of the modern condensed matter physics. They provide valuable information on one- and many-body excitation spectra, such as quasiparticle dispersions, lifetimes, and emergent low energy excitations. In addition, they are indispensable for understanding and interpreting transport and thermodynamic measurements and are the direct target of spectroscopic probes such as ARPES, x-ray, and neutron scattering~\citep{Damascelli_RMP2003,Damascelli_RMP2024,RICS_RMP2011}.  At the same time, obtaining high-fidelity spectral functions often remains computationally demanding.

This difficulty spans across a wide range of settings:  impurity solvers, such as numerical renormalization group (NRG) and quantum Monte Carlo (QMC) methods, as well as diagrammatic and exact diagonalization techniques can be accurate but become costly when scanning multi-parameter Hamiltonians or when repeated evaluations are needed for inference and parameter exploration. This computational bottleneck motivates machine-learning surrogate models~\cite{} that learn either the forward map (Hamiltonian parameters $\rightarrow$ spectral function $A(\omega)$) or the inverse map ($A(\omega)$ $\rightarrow$ parameters), with the goal of accelerating parameter exploration and enabling data-driven inference~\citep{Arsenault_PRB2014,Sturm_PRB2021,LCY_PRB2023}. In practice, the learned map is often highly structured: a small number of Hamiltonian parameters can induce nontrivial changes in peak locations, widths, and spectral weight redistribution across the full energy window. This makes spectral surrogates an appealing testbed for data-efficient regression methods that can exploit smoothness while maintaining good accuracy with limited training data and can generalize across the underlying parameter space.

A number of recent works have demonstrated that modern machine learning (ML) can accurately interpolate spectral data when sufficiently large training sets are available. The pioneering work at this front was done by~\cite{Arsenault_PRB2014} where they applied various machine learned representations to calculate spectral function for the Anderson Impurity model (AIM), a paradigmatic model of quantum impurity. Building on this, \cite{Sturm_PRB2021} built a high-fidelity large NRG database ($\sim$5×10$^5$ spectra) and compared deep neural network (DNN)-based model to kernel ridge regression approach, with DNN generally superior at scale. Beyond forward prediction, \cite{Miles_PRB2021} used variational autoencoders to learn interpretable latent variables from single-impurity Anderson model (SIAM) spectra. 
More recently, Lee, Carbone, and Yin (LCY) (referred to as LCY hereafter) in 2023 generated $\sim 1.3\times 10^{5}$ self-consistent Born approximation (SCBA) spectra for the $t$--$t'$--$t''$--$J$ model describing the motion of a single hole in a quantum antiferromagnet~\citep{LCY_PRB2023}. They showed that simple feed-forward neural networks can accurately learn both the forward map and inverse parameter inference, yielding substantial speedups relative to direct SCBA evaluation~\citep{LCY_PRB2023}. Related progress has also appeared in ARPES-oriented ML workflows and in data-driven treatments of spectral reconstruction tasks (e.g., analytic continuation, denoising etc), where training data can be limited and robustness is essential~\citep{Ekahana_2023,Damascelli_2025,Bian_ClusteringARPES_CommPhys2024, Kim_DenoisingARPES_Kim2021}.

At the same time, two practical issues repeatedly arise when moving from proof-of-principle demonstrations toward workflows for scientific discovery. First, many spectral surrogates still require large training sets, whereas in many applications high-quality spectra are expensive to simulate and only a modest number of computer model-based solver calls are feasible under tight compute budgets. Second, point predictions can be difficult to use in downstream tasks such as active learning, experimental design, or inverse problems, where it is helpful to know when the surrogate is operating within its training regime versus extrapolating~\citep{donatelli2025basic}; related surrogate and inference workflows that incorporate reliability and error-awareness are now routine in other computational science settings, including probabilistic projections~\citep{jantre2024probabilistic} and computer model system-identification~\citep{jantre2025data}. In spectroscopy ML, lightweight reliability indicators can be useful, especially when out-of-distribution inputs and systematic errors are possible, and they can help contextualize model predictions~\citep{Ghose_PRB2023, Carbone_JPCA2024}.

These considerations naturally motivate Gaussian-process (GP) surrogates for spectral-function modeling in low-dimensional Hamiltonian parameterizations. GP regression provides a flexible nonparametric surrogate that encodes smoothness through its kernel and is well known to be effective in small-data regimes~\citep{rasmussen2005-GP}. From a physics perspective, GP priors can be viewed as enforcing controlled smoothness assumptions on the parameter-to-spectrum map, while allowing the data to determine the relevant correlation lengthscales. Moreover, GP models connect naturally to sequential design and active-learning workflows~\citep{santner2003design,cohn1994active}, and GP-based calibration frameworks provide a principled way to relate simulator outputs to experimental data in the presence of model discrepancy~\citep{kennedy2001calib}. The main historical obstacle has been scalability: exact GP regression scales cubically with the number of training data samples. However, sparse variational GP methods using inducing variables and stochastic optimization make GP regression practical at the $10^{4}$--$10^{5}$ scale~\citep{titsias2009variational, hensman2013GPbigdata}. Deep kernel learning (DKL) combines neural feature extraction with GP-based predictions by learning a data-adapted kernel, and stochastic variational formulations enable DKL to scale to large data while retaining GP approaches’ nonparametric modeling advantages~\citep{wilson2016_dkl_svgp}.

In this work, we develop a data-efficient surrogate for the LCY's SCBA database of the $t$--$t'$--$t''$--$J$ model. We focus on the forward task of predicting the density of states $A(\omega)$ from the three-dimensional Hamiltonian parameter vector $(t',t'',J)$ (with $t$ setting the overall energy scale as in LCY). Using only a uniformly random 10\% subset of the available $\sim10^{5}$ spectra, we train a deep-kernel stochastic variational Gaussian process (DKL-SVGP) surrogate model and benchmark against feed-forward neural networks trained on the same reduced budget as well as on the full training dataset. We find that our GP surrogate consistently outperforms the FFNN trained on the same reduced dataset and, despite using only 10\% of the training spectra, attains spectrum-wise errors of the same order-of-magnitude as the full-data FFNN baseline, while maintaining comparable peak-height accuracy.
Peak locations are typically accurate, but we observe rare large deviations motivating an explicit peak-matching diagnostic protocol that substantially reduces these outliers and improves peak-location fidelity. 
 Overall, these results show that our GP-based surrogate offers a compelling accuracy-data tradeoff for spectral-function modeling in scarce-data regimes and provides a natural foundation for active learning and inverse modeling workflows. 

\section{Methodology}
\subsection{Overview of the Physical System}
We study the spectral function of the \tJmodel model~\citep{t-J-YinPRL1998,Manousakis_PRB2007}, which is an improved description of the $t-J$ model, originally proposed to describe the motion of a single hole in a quantum antiferromagnet~\citep{Marsiglio_PRB1991,Martinez_PRB1991,Manousakis_PRB1992,Yin_PRB1997,Manousakis_PhysLettA2007}. For completeness, we write the Hamiltonian in Eq.~\ref{eq:Hamiltonian} and briefly summarize how the density of states (DOS) data used for training are generated.
\begin{eqnarray}
    H &=& -\left(
    t \sum_{\langle i,j\rangle_{1},\sigma}
    + t' \sum_{\langle i,j\rangle_{2},\sigma}
    + t'' \sum_{\langle i,j\rangle_{3},\sigma}
    \right)
    \bigl(\widetilde{c}^{\dagger}_{i\sigma}\widetilde{c}_{j\sigma} + \mathrm{H.c.}\bigr) \nonumber \\
    && + J \sum_{\langle i,j\rangle_{1}} \mathbf{S}_{i}\cdot\mathbf{S}_{j}\,
\label{eq:Hamiltonian}
\end{eqnarray}
where $\widetilde{c}^{\dagger}_{i\sigma}$, $\widetilde{c}_{i\sigma}$ are constrained fermion creation and annihilation operators forbidding double occupancy; $\langle i,j\rangle_n$ denotes $n^{th}$ neighbor pairs, and $\mathbf{S}_i$ are spin operators.

The angle-resolved spectral function is
\begin{equation*}
    A(\mathbf{k},\omega) = -\frac{1}{\pi}\,\mathrm{Im}\,G(\mathbf{k},\omega)
\end{equation*}
and the DOS is obtained by summing over momenta,
\begin{equation*}
    A(\omega)=\frac{1}{K}\sum_{\mathbf{k}}A(\mathbf{k},\omega).
\end{equation*}

$A(\omega)$ is computed with the non-crossing self-consistent Born approximation (SCBA)~\citep{Schmitt_PRL1988, Manousakis_PRB1991}, which iteratively evaluates the hole Green's function dressed by magnon excitations until convergence. In practice, SCBA requires dense sampling in both the hole momentum $\mathbf{k}$ and magnon momentum $\mathbf{q}$; the dataset we use corresponds to converged calculations on the grids reported in LCY.

The training corpus consists of a pool of $\sim1.3\times10^{5}$ DOS curves $A(\omega)$ generated on a dense uniform grid of $(t',t'',J)$ values with 51 points in each direction (with $t$ fixed as the energy unit), i.e., $t'\in[-0.5,0.5]$, $t''\in[-0.5,0.5]$, and $J\in[0.2,1.0]$. For each parameter triplet $(t',t'',J)$, the SCBA output DOS is sampled on a uniform grid of $301$ energies. In this work we focus on the \emph{forward} surrogate modeling task
\begin{equation*}
    (t',t'',J)\;\longmapsto\; A(\omega)\in\mathbb{R}^{301}.
\end{equation*}
We train on a uniformly random $10\%$ subset of the LCY's training data, and we evaluate using the same validation and test data as provided in LCY.

\begin{figure*}
\centering
\includegraphics[width=\linewidth]{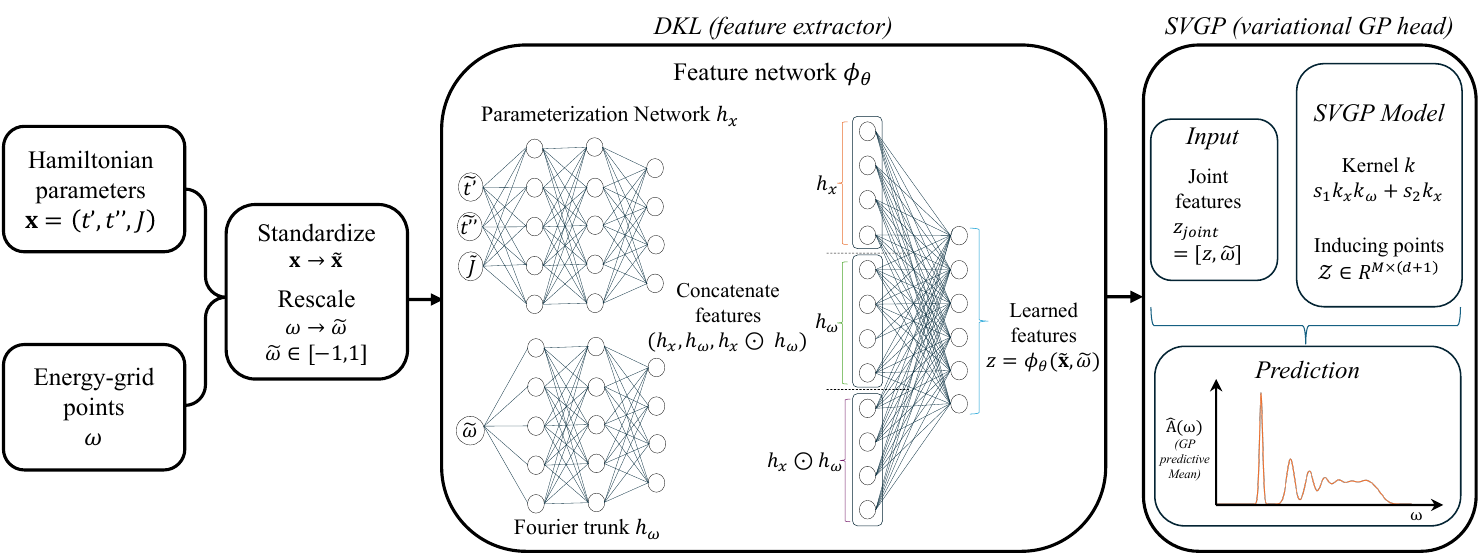}
\caption{Illustration of our deep kernel Gaussian process surrogate model trained with stochastic variational inference and applied to the forward problem of predicting a DOS given Hamiltonian parameters $x=(t', t'', J)$ and energy-grid points $\omega$. We first standardize $x\to \widetilde{x}$ and rescale $\omega \to \widetilde{\omega} \in [-1,1]$. Then, we pass $(\widetilde{x},\widetilde{\omega})$ through a feature network $(\phi_\theta$) consisting of parameterization network $h_x(\widetilde{x})$ and Fourier trunk $h_\omega(\widetilde{\omega})$, their outputs and element-wise interaction $(h_x\odot h_\omega)$ are concatenated and linearly mapped to form learned features $\mathbf{z} = \phi_\theta(\widetilde{x}, \widetilde{\omega})$. Finally, a stochastic variational GP is applied to the joint input $\mathbf{z}_{joint}=[\mathbf{z},\widetilde{\omega}]$ using a flexible kernel and $M$ inducing points, producing the surrogate  prediction which is GP mean -- $\widehat{A}(\omega)$.}
\label{fig:model-schematic}
\end{figure*}

\subsection{Machine Learning Models}
\paragraph{Feed-Forward Neural Network (FFNN).} 
As a baseline deterministic surrogate, we use a fully-connected ReLU multi-layer perceptron trained with the Adam optimizer to minimize mean-squared error (MSE) loss. For a target spectrum $y^{(i)}\in\mathbb{R}^{m}$ and prediction $\widehat y^{(i)}$, the per-spectrum and the overall loss is
\begin{equation*}
    L^{(i)}=\frac{1}{m}\sum_{j=1}^{m}\!\left(\widehat y^{(i)}_{j}-y^{(i)}_{j}\right)^{2},\qquad
    L=\frac{1}{|T|}\sum_{i\in T}L^{(i)}\,,
\end{equation*}
where $T$ represents the evaluation data. We tune hyperparameters on the validation set using grid search over the network architecture ($net$), batch size (bs), learning rate (lr), and learning-rate scheduler (lrs). For the reduced training set (size $\sim10^{4}$), the best-performing FFNN uses
$net=(3,\,32,\,64,\,128,\,256,\,301)$, $\mathrm{bs}=128$, $\mathrm{lr}=5\times10^{-3}$, and $\mathrm{lrs}=0.5$. For reference and to compare against a high-capacity baseline trained on the full dataset, we also consider the FFNN architecture used in LCY, $net=(3,\,170,\,340,\,510,\,680,\,850,\,1020,\,301)$ with $\mathrm{bs}=1024$, $\mathrm{lr}=10^{-3}$, and $\mathrm{lrs}=0.5$.

\vspace{2mm}
\paragraph{Gaussian Process.}
To obtain a data-efficient surrogate, we model the \emph{forward map} from Hamiltonian parameters to the DOS using Gaussian process (GP) regression. Rather than learning the full curve in one shot, we treat the DOS as a function of both the three-dimensional parameter vector and energy-grid points, i.e., $y(\mathbf{x},\omega)\approx A_{\mathbf{x}}(\omega)$ with $\mathbf{x}=(t',t'',J)$. In practice, each training spectrum sampled on a grid $\{\omega_j\}_{j=1}^{P}$ provides $P$ scalar observations, producing training pairs $\mathbf{s}=[\mathbf{x},\omega]\in\mathbb{R}^4$ with scalar targets $y(\mathbf{s})$.

A GP places a prior directly on the unknown function,
\begin{equation*}
    \begin{aligned}
      f(\mathbf{s}) &\sim \mathcal{GP}\,\bigl(0,\;k(\mathbf{s},\mathbf{s}')\bigr), \\
      y(\mathbf{s}) &= f(\mathbf{s}) + \epsilon, \;\; \epsilon\sim\mathcal{N}(0,\sigma_n^2),
    \end{aligned}
\end{equation*}
yielding a flexible nonparametric surrogate defined by the kernel $k(\cdot,\cdot)$. We standardize inputs $\mathbf{x}$ and rescale $\omega$ to a fixed interval, and also apply log-standardization to the DOS to reduce dynamic-range effects (see Appendix~\ref{app:gp_details}).

\begin{figure*}[t]
  \centering
  \includegraphics[width=\linewidth]{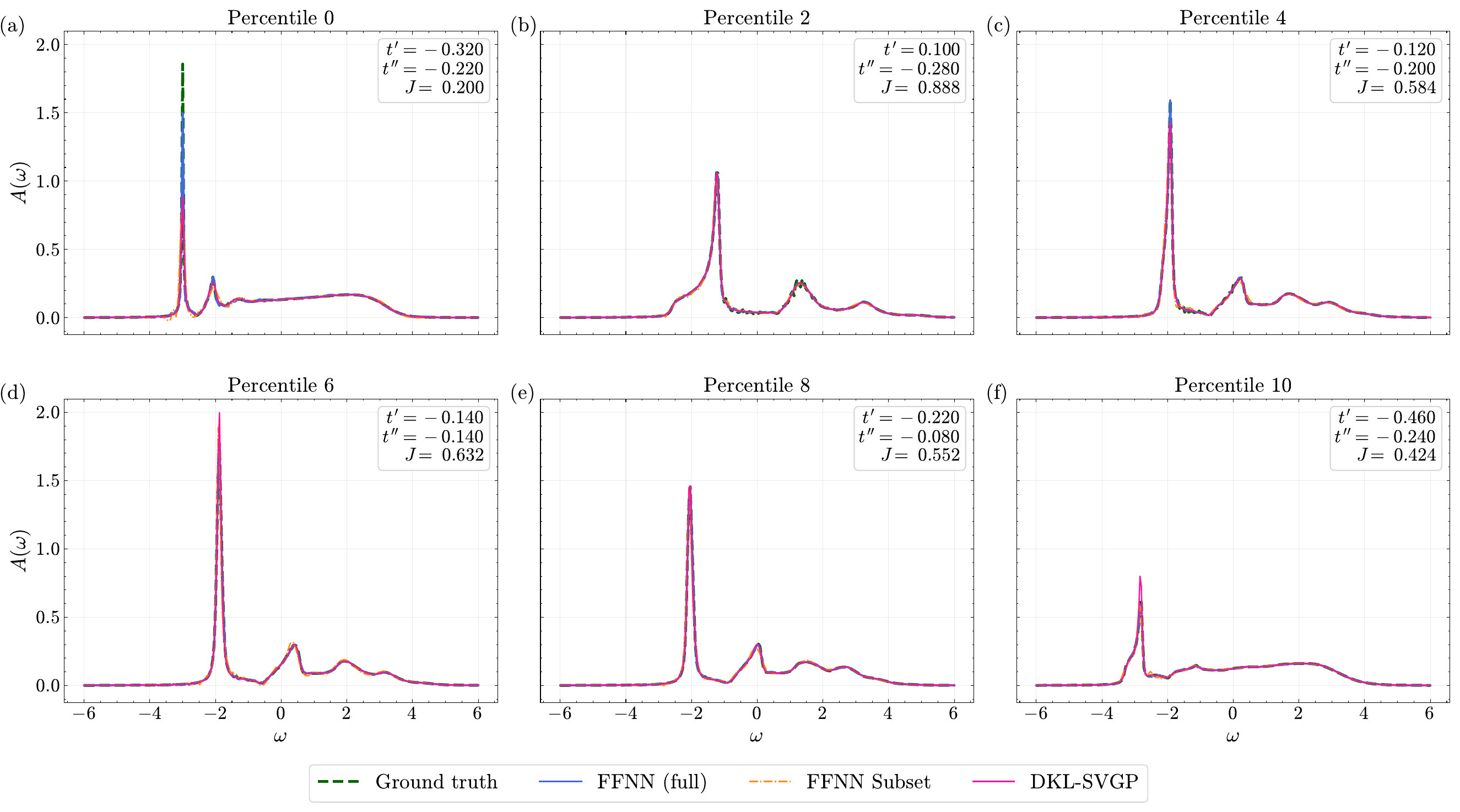}
  \caption{\textbf{Worst-tail diagnostic on the held-out test dataset.} Test spectra are ranked from worst to best by a reference error score (here, the FFNN row-RMSE over test dataset), and representative spectra at percentiles $\{0,2,4,6,8,10\}\%$ within this worst tail are shown. Each panel overlays the ground-truth DOS with predictions from the full-data FFNN baseline, the FFNN trained on a 10\% random subset (FFNN Subset), and our DKL-SVGP model trained on the same 10\% subset; the corresponding $(t',t'',J)$ values are annotated in each subplot.}
  \label{fig:worsttail}
\end{figure*}

A key practical issue is scalability: exact GP training scales as $\mathcal{O}(N^{3})$ in the number of training points $N$, and becomes prohibitive as the number of training pairs $N_{\text{pairs}}=N_{\text{spec}}\times P$ grows. We therefore use a sparse variational formulation with inducing variables, which makes GP learning practical at our dataset size~\citep{titsias2009variational,hensman2013GPbigdata}. To improve expressivity beyond a stationary kernel on raw inputs, we adopt deep kernel learning (DKL): a lightweight neural network feature map $\phi_\theta(\cdot)$ is learned jointly with the GP so that similarity is measured in a task-adapted representation rather than directly in $[\mathbf{x},\omega]$ space~\citep{wilson2016_dkl_svgp}. Concretely, we use an end-to-end DKL-SVGP surrogate model with an implicit kernel of the form $k(\mathbf{s},\mathbf{s}')=k_{\mathrm{base}}(\phi_\theta(\mathbf{s}),\phi_\theta(\mathbf{s}'))$, trained with minibatch stochastic optimization. Figure~\ref{fig:model-schematic} summarizes the model architecture and data flow.

After hyperparameter tuning, we use $M=1536$ inducing points and a Mat\'ern-$3/2$ kernel in the learned feature space, and we model the $\omega$-dependence with a flexible one-dimensional kernel; all architecture, kernel-construction, and training details (including preprocessing) are summarized in Appendix~\ref{app:gp_details}. 
In all GP experiments reported here, we train on the same uniformly random $10\%$ subset used for the reduced-data FFNN baseline and evaluate on the validation and test splits provided by LCY, enabling a direct comparison of data efficiency and predictive accuracy. We have reported deterministic predictions given by the GP predictive mean.

\subsection{Evaluation Protocol}
\label{sec:eval_protocol}
To compare surrogates, we report per-spectrum root-mean-square error (row-RMSE) and corresponding normalized variant (NRMSE) computed across the sampled energy-grid points $\{\omega_j\}_{j=1}^{P}$. For a ground-truth spectrum $\mathbf{y}_i=\{y_i(\omega_j)\}_{j=1}^{P}$ and prediction $\hat{\mathbf{y}}_i$, we define a small constant $\varepsilon$ for numerical stability. We summarize 
\begin{equation*}
    \begin{aligned}
        \mathrm{RMSE}_i &= \sqrt{\frac{1}{P}\sum_{j=1}^{P}\bigl(\widehat y_i(\omega_j)-y_i(\omega_j)\bigr)^2}, \\
    \mathrm{NRMSE}_i &= \frac{\mathrm{RMSE}_i}{\sqrt{\frac{1}{P}\sum_{j=1}^{P}y_i(\omega_j)^2}+\varepsilon},
    \end{aligned}
\end{equation*}
performance by the mean and median of $\mathrm{RMSE}_i$ and $\mathrm{NRMSE}_i$ over the train, validation, and test data splits provided by LCY, and report the results in Tables~\ref{tab:row_rmse_all} and~\ref{tab:nrmse_all}, respectively.

Additionally, we use a \emph{worst-tail percentile} diagnostic to stress-test models on difficult spectra. For a given split, we rank spectra from worst to best using a reference error score (here, the full-data FFNN baseline row-RMSE) and plot representative cases at percentiles $\{0,2,4,6,8,10\}\%$ within this worst tail (Fig.~\ref{fig:worsttail}). This diagnostic plot highlights failure modes such as misplaced secondary peaks and incorrect spectral-weight redistribution in predicted spectra.  

We also quantify peak-level fidelity by comparing the dominant peak height and its location between predictions and ground truth. For the ground-truth spectrum, we define the dominant peak height and location using the global argmax (argument of the maximum),
\begin{eqnarray*}
    & A_{\max,i} & =\max_{j} y_i(\omega_j), \\
    & \omega_{\max,i} & =\omega_{\arg\max_{j} y_i(\omega_j)}.
\end{eqnarray*}
For each model prediction $\widehat y_i(\omega)$, we report the corresponding dominant-peak quantities $\widehat A_{\max,i}$ and $\widehat\omega_{\max,i}$ defined analogously by argmax. In addition, to reduce sensitivity to rare cases where dominant and secondary peaks have comparable heights (so that the identity of the ``dominant" peak can swap), we also report a matched-peak diagnostic in which we match the ground-truth dominant peak to the nearest peak among the top-$K$ predicted peaks (we chose $K=5$ which was sufficient to mitigate the outliers in our case).
In addition to RMSE, we report the mean absolute error (MAE) and the Pearson correlation between predicted and true peak quantities $(A_{\max,i},\omega_{\max,i})$ to summarize absolute deviations and overall linear agreement (Fig.~\ref{fig:peaks_scatter}).



\section{Results and Discussion}
\label{sec:results}
Tables~\ref{tab:row_rmse_all} and~\ref{tab:nrmse_all} summarize per-spectrum errors across train/validation/test data splits from LCY. We observe two consistent trends.

\begin{table}[b]
\caption{Per-spectrum error (row-RMSE) across data splits. Values are mean and median (in parentheses) over spectra in each data split and are shown as $10^{-3}\!\times\!\text{row-RMSE}$ (lower is better). ``FFNN (full)'' is the full training data baseline from LCY; ``FFNN Subset'' and ``DKL-SVGP" are trained on the same uniformly randomly selected $10\%$ subset of full training data from LCY.}
\vspace{1mm}
\label{tab:row_rmse_all}
\footnotesize
\centering
\setlength{\tabcolsep}{6pt}
\begingroup
\def\arraystretch{1.25}
\begin{tabular}{l|ccc}
\hline
Data Split & FFNN (full) & DKL-SVGP & FFNN Subset \\[2pt]
\hline
Train     & 1.793 (1.528) & 4.581 (3.327) & 7.800 (6.129) \\
Validation       & 1.933 (1.595) & 4.612 (3.366) & 7.881 (6.242) \\
Test      & 1.928 (1.588) & 4.544 (3.307) & 7.831 (6.119) \\
\hline
\end{tabular}
\endgroup
\end{table}

First, under the same scarce-data training budget (a uniformly random $10\%$ subset of the LCY training set), the DKL-SVGP surrogate is substantially more accurate than the corresponding reduced-data FFNN baseline. On the test split, DKL-SVGP reduces the mean row-RMSE from $7.831\times10^{-3}$ (FFNN Subset) to $4.544\times10^{-3}$, i.e., a $\approx 42\%$ reduction, while the median row-RMSE drops from $6.119\times10^{-3}$ to $3.307\times10^{-3}$ (a $\approx 46\%$ reduction). A similar improvement is observed in the normalized metric: the mean test NRMSE decreases from $59.449\times10^{-3}$ to $33.685\times10^{-3}$ (a $\approx 43\%$ reduction) and the median test NRMSE reduces from $47.755\times10^{-3}$ to $25.571\times10^{-3}$ (a $\approx 46\%$ reduction). This indicates that the gain is not driven solely by a few high-amplitude spectra but persists after per-spectrum normalization. This behavior is aligned with the qualitative intuition emphasized in LCY’s forward-problem analysis: interpolation quality is most meaningfully judged by whether the surrogate preserves spectral morphology--peak locations, widths, and heights--even on the hardest examples in the test set. In that spirit, our worst-tail diagnostic (Fig.~\ref{fig:worsttail}) mirrors the LCY presentation of ``worst percentiles" and provides a stringent stress test focused on difficult spectra rather than average-case performance~\citep{LCY_PRB2023}. In these worst-tail panels, the DKL-SVGP predictions generally track the dominant peak structure and capture secondary features more reliably than the reduced-data FFNN, consistent with the quantitative advantage in Tables~\ref{tab:row_rmse_all}--\ref{tab:nrmse_all}.

\begin{table}[t]
\caption{Per-spectrum normalized error (NRMSE) across splits. Values are mean and median (in parentheses) over spectra in each split and are shown as $10^{-3}\!\times\!\text{NRMSE}$ (lower is better); divide by $10^3$ to recover raw NRMSE. ``FFNN (full)'' is the full training data baseline from LCY; ``FFNN Subset'' and ``DKL-SVGP'' are trained on the same uniformly random $10\%$ subset of the LCY training data~\cite{LCY_PRB2023}.}
\vspace{1mm}
\label{tab:nrmse_all}
\scriptsize
\centering
\setlength{\tabcolsep}{6pt}
\begingroup
\def\arraystretch{1.25}
\begin{tabular}{l|ccc}
\hline
Data Split & FFNN (full) & DKL-SVGP & FFNN Subset \\
\hline
Train & 13.655 (11.941) & 33.805 (25.786) & 59.151 (47.552) \\
Validation   & 14.714 (12.527) & 34.166 (25.902) & 59.935 (48.503) \\
Test  & 14.638 (12.453) & 33.685 (25.571) & 59.449 (47.755) \\
\hline
\end{tabular}
\endgroup
\end{table}

Second, while DKL-SVGP uses only $10\%$ of the training spectra, its errors remain within the same order-of-magnitude as the full-data FFNN baseline reported by LCY, though the full-data model is still more accurate. Concretely, on the test split the mean row-RMSE is $1.928\times10^{-3}$ for FFNN (full) versus $4.544\times10^{-3}$ for DKL-SVGP, i.e., the GP surrogate is within a factor of $\sim 2.4$ of the full-data network despite a $10\times$ ($1$ order-of-magnitude) smaller training set. Importantly, for all three models the error statistics are consistent across splits (train $\approx$ val $\approx$ test), suggesting that the reported differences reflect genuine generalization behavior rather than overfitting or data leakage. From a modeling perspective, this gap between DKL-SVGP and the full-data FFNN is expected: with enough data, a high-capacity deterministic network can fit the forward map extremely well. The key result here is that, in the limited-data regime where the reduced-data FFNN begins to lose fidelity, the GP-based surrogate recovers a large fraction of the full-data accuracy without requiring a comparably large training corpus. We also include two additional classical baselines in Appendix~\ref{app:additional_baselines}: kernel ridge regression (KRR), which tests how far a stationary kernel on the raw Hamiltonian parameters can go, and k-nearest neighbors (kNN), a simple distance-based baseline. We place both in Appendix~\ref{app:additional_baselines} to keep the main discussion focused on the data efficiency and spectrum-level fidelity of DKL-SVGP versus FFNN, while noting that both KRR and kNN perform substantially worse than DKL-SVGP on this task (test RMSE $\sim 15\times10^{-3}$ for KRR/kNN versus $\sim 4.5\times10^{-3}$ for DKL-SVGP).

\begin{figure*}[t]
  \centering
  \includegraphics[width=\linewidth]{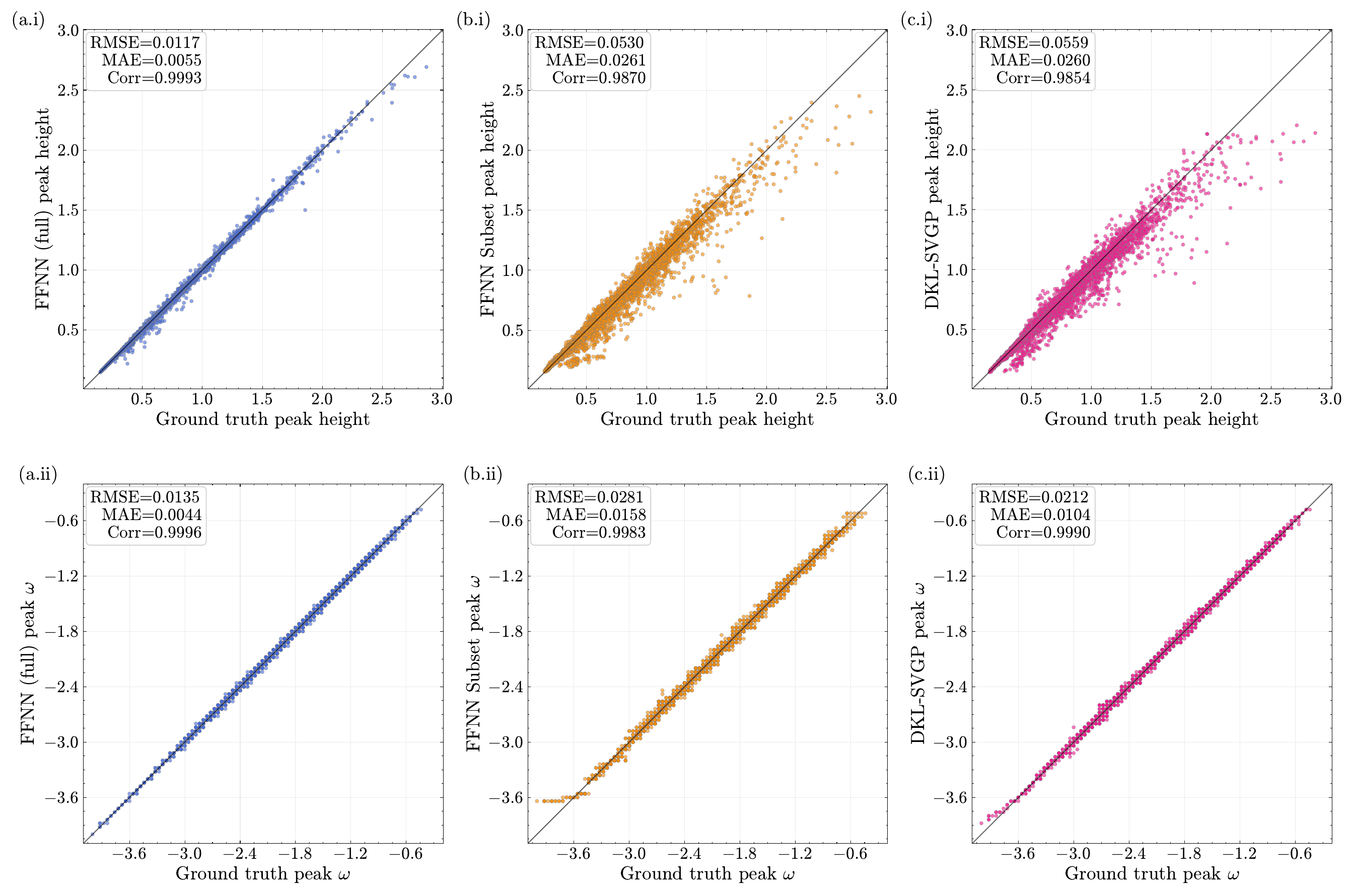}
  \caption{\textbf{Peak-level accuracy on the test dataset with top-5 peak matching.} Scatter plots compare predicted vs. ground-truth peak height (top row) and peak location (bottom row) for the full-data FFNN baseline, the FFNN trained on a uniformly random 10\% subset, and DKL-SVGP trained on the same 10\% subset. For each spectrum, peak errors are computed after matching the ground-truth dominant peak to the closest peak among the top-5 peaks of each model prediction. Each panel reports RMSE, MAE, and Pearson correlation with respect to the ground truth.}
  \label{fig:peaks_scatter_matched_top5}
\end{figure*}

Finally, many downstream uses of spectral surrogates depend on peak-level physics; for example, peak positions encode information about excitation energies and band dispersion, whereas peak heights are related to the quasiparticle weights. Therefore, we complement spectrum-wise errors with peak diagnostics (Fig.~\ref{fig:peaks_scatter_matched_top5}). Scatter plots of the dominant peak height $A_{\max}$ and corresponding location $\omega_{\max}$ assess how well each surrogate preserves salient low-energy structure, in the same spirit as LCY’s emphasis on reproducing peak positions and heights when discussing challenging spectra~\citep{LCY_PRB2023}. To this end, we compare predicted versus ground-truth dominant peak height $A_{\max}$ (top row) and peak location $\omega_{\max}$ (bottom row), using the matched-peak definition in Section~\ref{sec:eval_protocol} to reduce sensitivity to rare cases of dominant/secondary (top-$K$ matching with $K=5$). Under this diagnostic, peak-height accuracy is broadly comparable between the two reduced-data models: the FFNN trained on a 10\% subset attains RMSE $\approx 0.053$ and MAE $\approx 0.026$, while DKL-SVGP yields RMSE $\approx 0.056$ and MAE $\approx 0.026$ (both with strong correlations $\gtrsim 0.98$). Peak-location accuracy is better for DKL-SVGP: it attains RMSE $\approx 0.021$ compared to $\approx 0.028$ for the reduced-data FFNN, MAE $\approx 0.010$ versus $\approx 0.016$, with correlations $\gtrsim 0.99$ for both. For completeness, we report diagnostic results under strict argmax definition and related outlier diagnostics in Appendix~\ref{sec:peak_appendix}, where the largest peak-location deviations are shown to arise primarily from occasional peak swapping rather than a complete loss of spectral fidelity.

Overall, these results highlight that the advantages of DKL-SVGP lie in reducing spectrum-wise error under limited data (Tables~\ref{tab:row_rmse_all}--\ref{tab:nrmse_all}), improving overall spectral curve fidelity (Fig.~\ref{fig:worsttail}), and recovering peak height and location more accurately under the matched-peak definition. This interpretation supports the main practical takeaway: replacing a deterministic neural surrogate with a DKL-SVGP model can yield large accuracy gains in scarce-data regimes without increasing the training-data budget. In all experiments reported here, we use deterministic predictions given by the GP predictive mean.

\section{Conclusion}
\label{sec:conclusion}
We developed a data-efficient surrogate for SCBA density-of-states spectra of the $t$--$t'$--$t''$--$J$ model by combining deep kernel learning with sparse variational Gaussian process inference. Using only a uniformly random $10\%$ subset of LCY's training split for our surrogate modeling, the proposed DKL-SVGP model consistently outperforms a feed-forward neural network trained on the same subset, reducing the mean test row-RMSE by $\sim42\%$ and the mean test NRMSE by $\sim43\%$. While the full-data FFNN baseline remains the most accurate overall, DKL-SVGP trained on only $10\%$ of the data achieves the same order-of-magnitude spectrum-wise errors and maintains consistent generalization across the LCY train/validation/test data splits. Qualitatively, the model also yields more faithful full-spectrum reconstructions on difficult worst-tail examples, where errors are often driven by secondary-peak structure and spectral-weight redistribution. Peak-level diagnostics are broadly consistent with this trend, where our model and reduced-data FFNN perform comparably on peak-height metrics, and our model improves in peak-location metrics under the matched-peak evaluation. Taken together, these results demonstrate that GP-based surrogates provide a strong accuracy--data tradeoff for spectral modeling in scarce-data regimes.

Several extensions are natural. First, improved data-selection strategies (e.g., space-filling or physics-informed sampling near sharp spectral changes) could further reduce training requirements beyond uniform subsampling. Second, extending the surrogate to inverse parameter inference would enable end-to-end workflows for parameter calibration against spectral data. Finally, structured multi-output modeling of the full spectrum (rather than scalarized training pairs), as well as extensions to momentum-resolved spectra $A(\mathbf{k},\omega)$, are promising directions toward reliable and sample-efficient spectral solvers for broader condensed-matter workflows. 

\section{Acknowledgement}
\label{sec:acknowledgement}

S. Jantre and N. Aryal are supported by the Early Career Research Program funding from the Basic Energy Sciences (BES) program of the United States (US) Department of Energy's (DOE) Office of Science, under Award No. DE-SCL0000027.
W. Yin is supported by the funding from the BES program's Materials Sciences and Engineering Division of the US DOE's Office of Science, under Contract No. DE-SC0012704.
N. Urban is supported by the funding from the Advanced Scientific Computing Research program of the US DOE's Office of Science, under Contract No. DE-SC0012704.
S. Jantre and N. Aryal also acknowledge the support of the Laboratory Directed Research and Development Grant (LDRD \# 24-039) from Brookhaven National Laboratory at the initial stage of the project.


\appendix

\section{DKL-SVGP Surrogate Details}
\label{app:gp_details}
This appendix summarizes the preprocessing, model construction, and training setup used for the DKL-SVGP surrogate reported in the main text.

\subsection{Training Pairs and Preprocessing}
Each spectrum $A(\omega)$ is provided on a fixed grid $\{\omega_j\}_{j=1}^{P}$. We form scalar regression pairs
\begin{equation*}
    \mathbf{s}_{ij} = [\mathbf{x}_i,\ \omega_j]\in\mathbb{R}^4,\qquad
    y_{ij} = A_{\mathbf{x}_i}(\omega_j),
\end{equation*}
Here $N$ is the number of spectra and $P$ is the number of energy-grid points per spectrum. Hence, a dataset of $N$ spectra yields $N \cdot P$ scalar observations.

\paragraph{Input standardization.}
We standardize Hamiltonian parameters dimension-wise using training dataset's mean and standard deviation:
\begin{equation*}
    \widetilde{\mathbf{x}} = (\mathbf{x}-\boldsymbol{\mu}_x) \oslash (\boldsymbol{\sigma}_x + 10^{-8}),
\end{equation*}
where $\oslash$ denotes an element-wise vector division. $\boldsymbol{\mu}_x, \boldsymbol{\sigma}_x \in \mathbb{R}^3$ are computed over the training dataset. Next, we rescale the energy grid to a fixed interval via an affine transformation
\begin{equation*}
    \widetilde{\omega} = 2\,\frac{\omega-\omega_{\min}}{\omega_{\max}-\omega_{\min}+10^{-12}}-1 \in [-1,1].
\end{equation*}

\paragraph{Log-standardization of the DOS.}
To reduce the dynamic-range effect, we apply a per-$\omega$ log transform followed by standardization:
\begin{equation*}
    z(\omega) = \log\,(\,\max(A(\omega),\,0)+\varepsilon),\qquad \varepsilon=10^{-3},
\end{equation*}
and compute training-dataset mean and standard deviation at each grid point, $\mu_z(\omega_j)$ and $\sigma_z(\omega_j)$. The GP is trained on
\begin{equation*}
    \widetilde{y}(\omega_j)=\frac{z(\omega_j)-\mu_z(\omega_j)}{\sigma_z(\omega_j)+10^{-8}}.
\end{equation*}
At inference time, predictions are mapped back using the inverse transformation 
$$A(\omega_j)=\exp\!\big(\,\widetilde{y}(\omega_j)\,\sigma_z(\omega_j)+\mu_z(\omega_j)\big)-\varepsilon$$

\subsection{Deep Feature Network}
We use a lightweight feature network $\phi_{\theta}$ ($\theta$ denote the network parameters) to learn a task-adapted representation before applying the GP regression. Given $\mathbf{s}=[\widetilde{\mathbf{x}},\widetilde{\omega}]$, the network produces a learned feature vector $\mathbf{z} = \phi_\theta(\mathbf{s}) \in \mathbb{R}^{d}$ with $d=128$. The network architecture is intentionally small:
\begin{itemize}
    \item A two-layer MLP $h_x(\widetilde{\mathbf{x}})$ for the three-dimensional parameter input.
    \item A learnable Fourier trunk $h_\omega(\widetilde{\omega})$ built from (i) fixed Fourier features $\{\sin(2\pi \kappa \,\widetilde{\omega}),\cos(2\pi \kappa \,\widetilde{\omega})\}_{\kappa=1}^{n_f}$ with $n_f=16$, and (ii) learned-frequency features $\{\sin(2\pi \nu_\ell\,\widetilde{\omega}),\cos(2\pi \nu_\ell\,\widetilde{\omega})\}_{\ell=1}^{n_\ell}$ with $n_\ell=8$, where $\nu_\ell=\mathrm{softplus}(\eta_\ell)>0$ are trainable frequencies.
    \item A multiplicative interaction channel $h_x(\widetilde{\mathbf{x}})\odot h_\omega(\widetilde{\omega})$. Here, $\odot$ is element-wise vector product.
    \item A linear projection to $d$ dimensions, followed by $\ell_2$ normalization (to stabilize kernel lengthscales).
\end{itemize}

\subsection{Kernel Construction}
As the final step of the DKL-SVGP surrogate, the GP is applied to an augmented representation
\begin{equation*}
    \mathbf{z}_{\mathrm{joint}} = [\phi_\theta(\mathbf{s}),\, \widetilde{\omega}] \in \mathbb{R}^{d+1},
\end{equation*}
so that the model can (i) compare inputs in learned feature space and (ii) retain an explicit one-dimensional coordinate for $\widetilde{\omega}$.

We use a flexible kernel family of the form
\begin{equation}
\begin{aligned}
    k(\mathbf{z}_{\mathrm{joint}},\mathbf{z}'_{\mathrm{joint}}) & = s_1\,k_x(\phi_\theta(\mathbf{s}),\phi_\theta(\mathbf{s}'))\,k_\omega(\widetilde{\omega},\widetilde{\omega}') \\
    & \qquad + s_2\,k_x(\phi_\theta(\mathbf{s}),\phi_\theta(\mathbf{s}')),
\end{aligned}
\label{eq:flex_kernel_appendix}
\end{equation}
where $s_1,s_2>0$ are learned amplitudes that control the relative strength of the coupled $(\mathbf{x},\omega)$ interaction kernel and the feature-only kernel.

\vspace{2mm}
\paragraph{Learned-feature kernel $k_x$.}
The kernel $k_x$ is Mat\'ern-$3/2$ with automatic relevance determination (ARD) lengthscales in the $d$-dimensional feature space.

\vspace{1.5mm}
\paragraph{Energy-grid point kernel $k_\omega$.}
We model the energy dependence with a one-dimensional Spectral Mixture (SM) kernel with $Q=16$ mixtures, which is expressive enough to capture multi-scale and quasi-periodic structure in $\omega$. The SM parameters are initialized with evenly spaced mixture means and moderate initial scales.

\subsection{Sparse Variational GP (SVGP)}
Because the scalarized dataset contains $N \cdot P$ training pairs, exact GP training is impractical. We therefore use a sparse variational GP with inducing variables for GP inference. Let $\mathbf{u}=f(\mathbf{Z})$ denote the latent GP function values at inducing inputs $\mathbf{Z} = \{\mathbf{z}_m\}_{m=1}^M$, where $\mathbf{Z} \in \mathbb{R}^{M\times(d+1)}$ and $M=1536$. We optimize a Gaussian variational distribution $q(\mathbf{u})=\mathcal{N}(\mathbf{m},\Sigma)$, with $\mathbf{m}$ and $\Sigma$ denoting mean vector and covariance matrix respectively, by maximizing the standard variational evidence lower bound (ELBO) via mini-batching over training pairs. Inducing locations are initialized by $k$-means clustering on a random subset of up to $5\times 10^4$ training pairs and are learned during training.

\subsection{Training Configurations}
For DKL-SVGP model results, we train our model end-to-end with minibatch stochastic optimization using:
\begin{itemize}
    \item Inducing points: $M=1536$; learn inducing locations; Cholesky variational distribution (full-rank Gaussian $q(\mathbf{u})$ parametrized via a Cholesky factor); whitened parameterization (reparameterization of inducing variables for better conditioning).
    \item Precision: float64.
    \item Batch size: 16384 scalar pairs.
    \item Optimizers: Natural Gradient Descent (NGD) for variational parameters (learning rate $2\times 10^{-4}$) and Adam for the rest (learning rate $8\times 10^{-4}$, with weight decay of $10^{-5}$ on the feature network).
    \item NGD schedule: delay 30 epochs, linear ramp over next 30 epochs.
    \item Stabilization: Cholesky jitter $5\times 10^{-3}$; gradient clipping (norm 1.0) for feature and likelihood parameters; clamping of unconstrained kernel parameters (e.g., lengthscales/scales) and inducing locations to prevent numerical overflow.
    \item Likelihood noise: Gaussian likelihood with noise variance $\sigma_n^2$ clamped to $[\sigma^2_{\min},\sigma^2_{\max}(e)]$ where $\sigma^2_{\min}=10^{-5}$ and $\sigma^2_{\max}(e)$ is annealed from 0.25 to 0.03 over the first 35 epochs.
    \item Warmup: 50 steps optimizing likelihood noise only.
    \item Freeze schedule: SM kernel parameters for $k_\omega$ are frozen for the first 45 epochs, then unfrozen.
    \item Early stopping: patience period of 40 epochs based on validation ELBO.
\end{itemize}

\subsection{Model Prediction}
Given a new input sample $(\mathbf{x},\omega)$, we apply the same preprocessing to obtain $(\widetilde{\mathbf{x}},\widetilde{\omega})$, evaluate the variational GP predictive distribution in the transformed space, and map predictions back to $A(\omega)$ via inverse log-standardization. The GP predictive mean yields the surrogate spectrum prediction.

\section{Additional Baseline Models}
\label{app:additional_baselines}
\paragraph{Kernel Ridge Regression (KRR):}
As an additional classical kernel-based baseline--complementary to the neural-network surrogate and our deep-kernel GP--we consider kernel ridge regression (KRR) for learning the forward map $(t',t'',J)\mapsto A(\omega)$ using the same 10\% subset of the LCY training dataset, using the sci-kit learn implementation~\citep{scikit-learn}. Like GP regression, KRR learns a nonlinear map via a positive-definite kernel, but it yields a deterministic predictor obtained by solving a regularized least-squares problem in a reproducing kernel Hilbert space. In our setting, KRR serves as a simple reference point for how well a stationary kernel on the raw Hamiltonian parameters can capture the relevant structure of the spectra.

Concretely, let $\{\mathbf{x}_i,\mathbf{y}_i\}_{i=1}^{N}$ denote training data with $\mathbf{x}_i=(t'_i,t''_i,J_i)\in\mathbb{R}^3$ and $\mathbf{y}_i\in\mathbb{R}^{P}$ the DOS sampled on the fixed grid $\{\omega_j\}_{j=1}^{P}$. We use a multi-output (vector-valued) KRR model with a shared kernel over $\mathbf{x}$, i.e.,
\begin{equation*}
    \begin{aligned}
        & \min_{f\in\mathcal{H}_k}\; \sum_{i=1}^{N}\bigl\| \mathbf{y}_i - f(\mathbf{x}_i)\bigr\|_2^2 +\lambda \|f\|_{\mathcal{H}_k}^2, \\
        & f(\mathbf{x}) = \sum_{i=1}^{N} \boldsymbol{\alpha}_i\, k(\mathbf{x},\mathbf{x}_i),
    \end{aligned}
\end{equation*}
where $\lambda>0$ is the ridge regularization parameter and $\boldsymbol{\alpha}_i\in\mathbb{R}^{P}$ are learned coefficients.

We adopt Gaussian radial-basis-function (RBF) kernel
\begin{equation*}
    k(\mathbf{x},\mathbf{x}') = \exp\!\left(-\gamma \|\mathbf{x}-\mathbf{x}'\|^2\right),
\end{equation*}
with width parameter $\gamma>0$ (equivalently $\gamma = 1/(2\ell^2)$ in terms of a lengthscale $\ell$). We tune $(\gamma,\lambda)$ by grid search on the validation split using the same spectrum-wise error metric as for the FFNN and DKL-SVGP surrogates. 
Over the explored range of hyperparameters, the best KRR model attains a test-set RMSE of $15.5\times 10^{-3}$ at $\lambda=10^{-6}$ and $\gamma=1.7$. This is roughly four times larger than the DKL-SVGP RMSE, suggesting that a stationary RBF kernel captures only coarse smooth trends in $A(\omega)$ as a function of $(t',t'',J)$ and struggles with sharper, parameter-dependent spectral rearrangements that benefit from learned representations.

\paragraph{k-Nearest Neighbors (kNN):}
Following LCY, we also include a $k$-nearest neighbors (kNN) regression baseline. kNN is a non-parametric, lazy learning method~\citep{scikit-learn,Biau_2015}: given a training set of input--output pairs $\{(\mathbf{x}_i,\mathbf{y}_i)\}_{i=1}^{N}$, the kNN prediction, $\hat{\mathbf{y}}(\mathbf{x})$, at a query point $\mathbf{x}$ is formed by a weighted average over the outputs of the $k$ closest training inputs.
\begin{equation*}
\hat{\mathbf{y}}(\mathbf{x})=\sum_{n \in \mathcal{N}_k(\mathbf{x})} w_n(\mathbf{x})\,\mathbf{y}_n ,
\end{equation*}
where $\mathcal{N}_k(\mathbf{x})$ denotes the index set of the $k$ nearest neighbors of $\mathbf{x}$ under a Minkowski distance
\begin{equation*}
d_n(\mathbf{x})=\lVert \mathbf{x}_n-\mathbf{x}\rVert_p
=\left(\sum_{j}\lvert x_{nj}-x_{j}\rvert^{p}\right)^{1/p}.
\end{equation*}
We use inverse-distance weighting, so closer neighbors contribute more strongly,
\begin{equation*}
w_n(\mathbf{x})=\frac{d_n(\mathbf{x})^{-1}}{\sum_{m\in \mathcal{N}_k(\mathbf{x})} d_m(\mathbf{x})^{-1}} \enskip \cdot
\end{equation*}
With the convention that if there exists a neighbor $n\in \mathcal{N}_k (\mathbf{x})$ with $d_n(\mathbf{x})=0$, we assign that neighbor weight 1 and the rest 0. By construction $\sum_{n \in \mathcal{N}_k (\mathbf{x})}w_n(\mathbf{x}) =1$ for all $\mathbf{x}$. We tune the hyperparameters $(k,p)$ by grid search on the LCY's validation data split using the same spectrum-wise error metric as for the FFNN and DKL-SVGP surrogates. The best kNN model yields an RMSE of $15.3\times 10^{-3}$ for $p=1.75$ and $k=4$, which is close to the RMSE obtained with the KRR model and considerably (almost $4\times$) worse than the DKL-SVGP RMSE.

\section{Peak-level Analysis Continued}
\label{sec:peak_appendix}
\begin{figure*}[t]
  \centering
  \includegraphics[width=\linewidth]{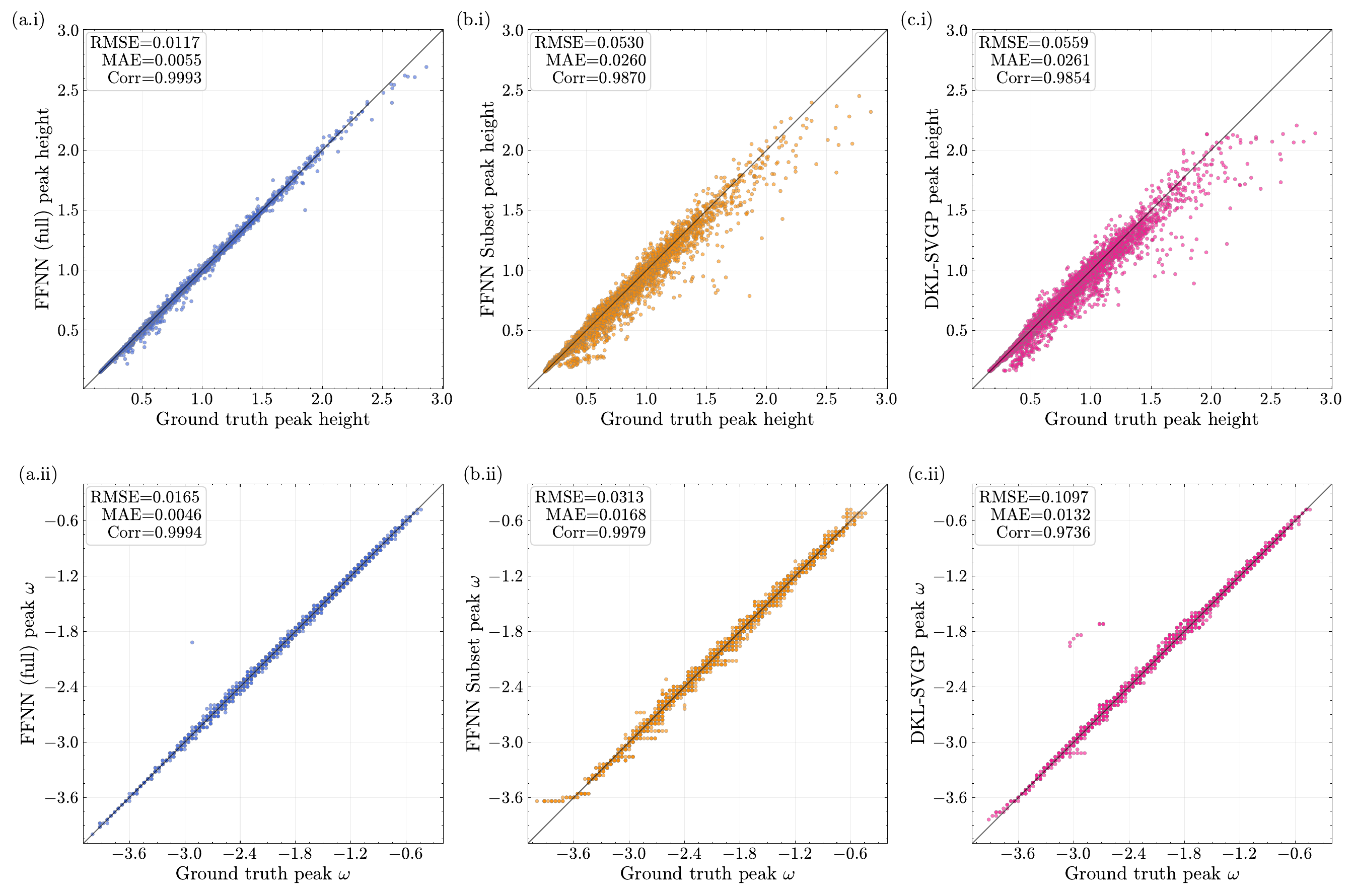}
  \caption{\textbf{Peak-level accuracy on the test dataset.} Scatter plots compare predicted vs. ground-truth (top row) dominant peak height $A_{\max}$ and (bottom row) peak location $\omega_{\max}$ for the full-data FFNN baseline, the FFNN trained on a 10\% random subset, and DKL-SVGP trained on the same 10\% subset. Each panel reports RMSE, MAE, and correlation of predictions with respect to the ground truth. We use consistent axis limits across panels for easier visual comparison; see Fig.~\ref{fig:peaks_scatter_extended} for an extended version showing additional outliers in DKL-SVGP peak location predictions.}
  \label{fig:peaks_scatter}
\end{figure*}

\begin{figure*}[t]
  \centering
  \includegraphics[width=\linewidth]{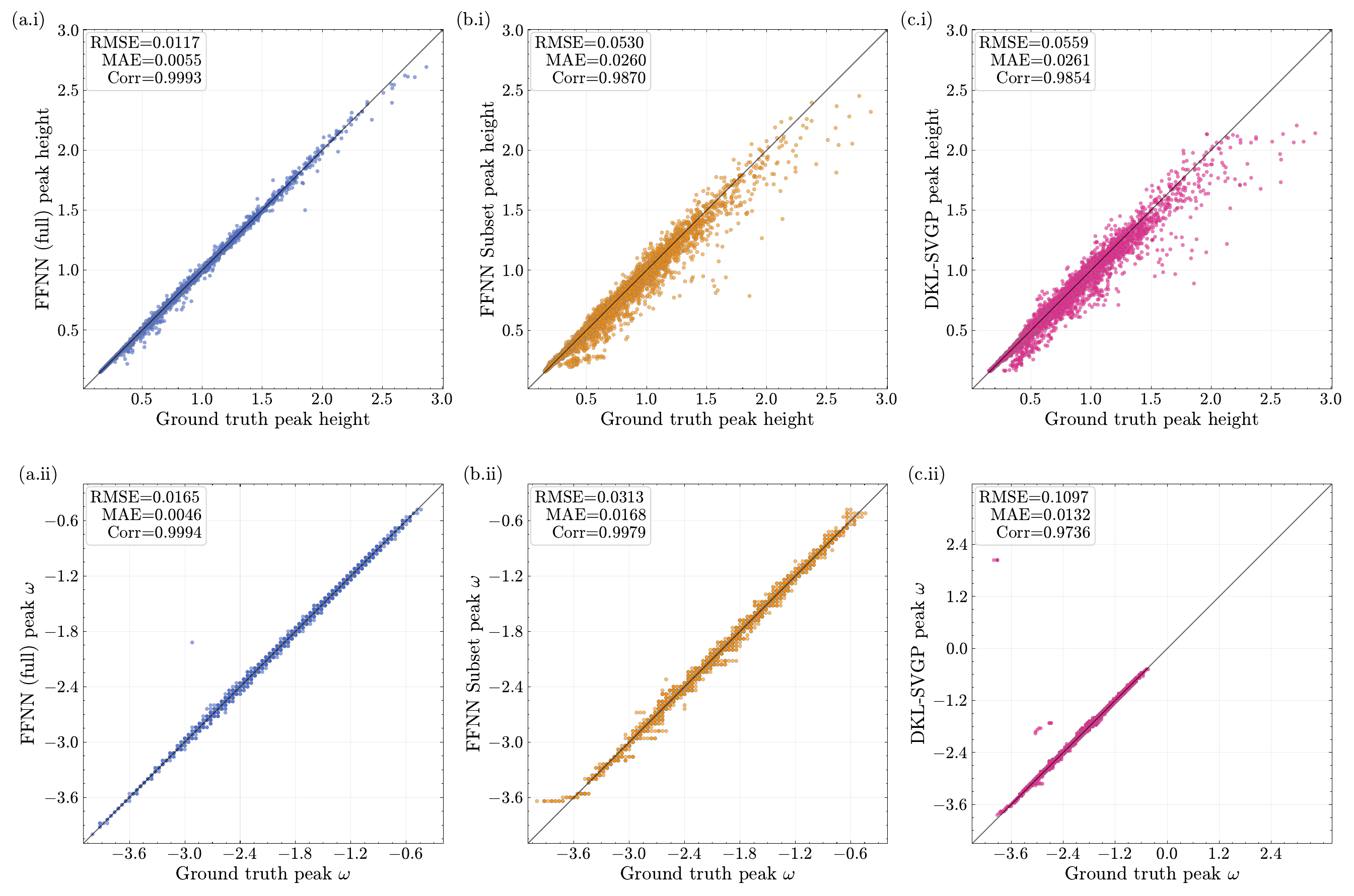}
  \caption{\textbf{Peak-level accuracy on the test dataset.} Scatter plots compare predicted vs. ground-truth (top row) dominant peak height $A_{\max}$ and (bottom row) peak location $\omega_{\max}$ for the full-data FFNN baseline, the FFNN trained on a 10\% random subset, and DKL-SVGP trained on the same 10\% subset. Each panel reports RMSE, MAE, and correlation of predictions with respect to the ground truth. Compared to Fig.~\ref{fig:peaks_scatter}, we extend the axis limits in the DKL-SVGP peak-location panel (c.ii) to highlight extreme outliers.}
  \label{fig:peaks_scatter_extended}
\end{figure*}

This appendix provides additional peak-level diagnostic plots under the \emph{strict argmax} peak definition. While the spectrum-wise metrics (Tables~\ref{tab:row_rmse_all}--\ref{tab:nrmse_all}) and worst-tail examples (Fig.~\ref{fig:worsttail}) indicate improved overall spectral fidelity for DKL-SVGP in the reduced-data regime, the strict argmax peak-location metric can exhibit a small number of pronounced outliers. We therefore provide two complementary views of the peak scatter plots: Fig.~\ref{fig:peaks_scatter} uses consistent axis limits across models for easy comparison (but necessarily clips extreme outliers), while Fig.~\ref{fig:peaks_scatter_extended} expands the axis limits to make the largest $\omega_{\max}$ discrepancies visible. 

As provided in Figs.~\ref{fig:peaks_scatter} and \ref{fig:peaks_scatter_extended}, we find that the peak-height accuracy is broadly comparable between the two reduced-data models: the FFNN trained on a 10\% subset attains RMSE $\approx 0.053$ and MAE $\approx 0.026$, while DKL-SVGP yields RMSE $\approx 0.056$ and MAE $\approx 0.026$ (both with correlations $\gtrsim 0.98$). In contrast, peak-location accuracy shows mixed behavior. While DKL-SVGP achieves a smaller MAE than the reduced-data FFNN (MAE $\approx 0.013$ vs.\ $0.017$), its RMSE is substantially larger and the correlation is lower (DKL-SVGP: RMSE $\approx 0.110$, corr.\ $\approx 0.974$; FFNN subset: RMSE $\approx 0.031$, corr.\ $\approx 0.998$), consistent with a small number of extreme outliers dominating the squared error.

\begin{figure*}[t]
  \centering
  \includegraphics[width=\linewidth]{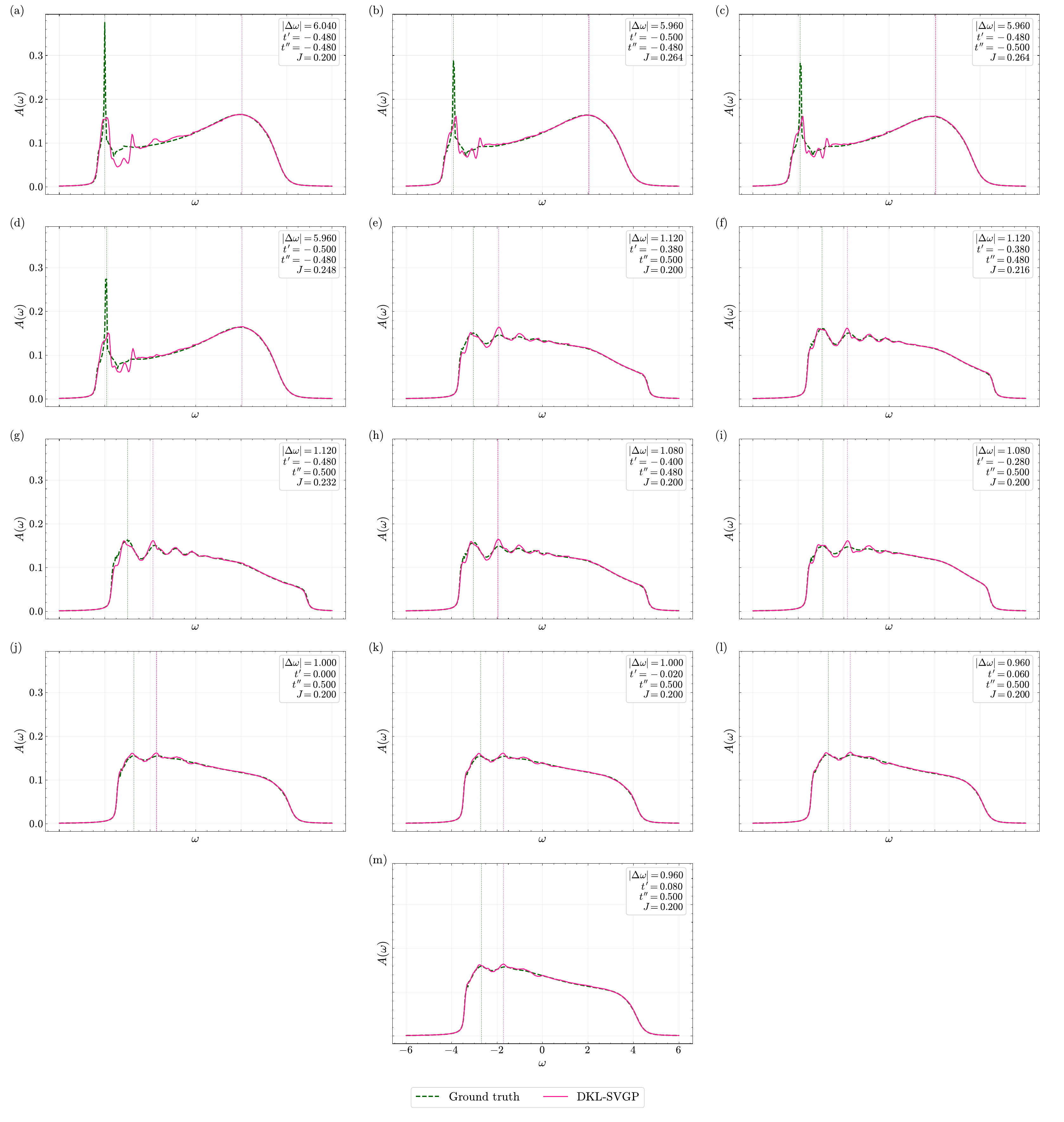}
  \caption{\textbf{Outlier peak-location cases (DKL-SVGP vs ground truth).} Diagnostic spectra for the 13 test-set outliers that deviate most noticeably in the peak-location scatter plot (Fig.~\ref{fig:peaks_scatter_extended}c.ii). Each panel overlays the ground-truth DOS and the DKL-SVGP prediction over the full energy range. Vertical dashed lines mark the dominant peak locations for the ground truth and for the DKL-SVGP prediction. Panels are annotated with the corresponding $(t',t'',J)$ values and the peak-location discrepancy $|\Delta\omega|$.}
  \label{fig:peakloc_outliers_spectral_predictions}
\end{figure*}

 To interpret these peak-location outliers, we next examine representative spectral overlays in Fig.~\ref{fig:peakloc_outliers_spectral_predictions} for the most prominent peak-location failures. Specifically, we select $13$ test-set cases with large peak-location discrepancy $|\Delta\omega| = |\omega_{\max}-\widehat{\omega}_{\max}| \ge 0.5$, and plot the full DOS for ground truth and DKL-SVGP prediction, with vertical dashed lines marking the respective argmax peak locations. These overlays show that many extreme $\omega_{\max}$ errors arise in spectra with competing peaks of similar height, where small relative-amplitude shifts can swap which nearby local maximum is labeled ``dominant," producing a large $\widehat{\omega}_{\max}$ shift despite otherwise reasonable spectral-shape agreement.

\clearpage

\end{document}